\documentclass[lettersize,journal]{IEEEtran}
\usepackage{amsmath,amsfonts}
\usepackage{algorithmic}
\usepackage{algorithm}
\usepackage{array}
\usepackage[caption=false,font=normalsize,labelfont=sf,textfont=sf]{subfig}
\usepackage{textcomp}
\usepackage{stfloats}
\usepackage{url}
\usepackage{verbatim}
\usepackage{graphicx}
\usepackage{cite}
\begin{document}

\title{Research on Train Shunting Impedance Based on Transmission Line Theory}

\author{Yinchao Dong and Linhai Zhao}

\markboth{Journal of \LaTeX\ Class Files,~Vol.~14, No.~8, August~2021}%
{Shell \MakeLowercase{\textit{et al.}}: A Sample Article Using IEEEtran.cls for IEEE Journals}


\maketitle

\begin{abstract}
At present, the shunting process of train to track circuit is usually studied by taking the shunting resistance of the first wheel set of train as the equivalent model, which ignores the shunting effect of other wheel sets and cannot study the fault conditions such as "pool shunting". Especially for the jointless track circuit (JTC), the compensation capacitors connected in parallel on the rail line will cause more complex train shunting process. To solve this problem, based on the transmission line theory, starting from the shunting resistance of each wheel set and considering the leakage of track bet, this paper established as six-terminal network model of train shunting impedance, indirectly verified the correctness of the model by using the working principle of track circuit reader (TCR), and focused on the influence of wheel set shunting resistance, compensation capacitors and rail line parameters on train shunting impedance. Finally, based on the calculation of the structural importance of train wheel set, a simplified model of train shunting impedance was constructed, providing theoretical support for the further study of fault conditions such as "poor shunting".
\end{abstract}


\section{Introduction}
\IEEEPARstart{A}{t} present, China's high-speed rail generally uses ZPW-2000 series of jointless track circuit (JTC), which uniformly installs compensation capacitors to ensure the effective transmission distance of the signal, but this makes the shunting process between train wheel sets and rail line more complex. As the operation of the train, the capacitive path formed by each compensation capacitor and the resistive path formed by each wheel set are interleaved and changed. Obviously, the current mathematical model that only equates the first wheel set resistance to the shunting process can no longer fully reflect this change process. Therefore, based on the train formation, with the whole train wheel sets as the research object, it is of great significance to study the apparent inlet impedance of the first wheel set to the end of the train, i.e., the train shunting impedance.

Currently, the first train wheel set resistance is stilled used as the equivalent of train shunting process\cite{zhao2009simulation, oukhellou2010fault, cherfi2012partially, cui2021fault}. This equivalent model is suitable for fault diagnosis of equipment connected in rail line such as compensation capacitors and tuning units based on train operation data\cite{zhao2011method, zhao2017line, zhao2013fault}, but for the study of other problems when train occupies the rail line, such as "poor shunting", is not applicable. In related studies, Jiawei Fu\cite{fu2018research} considered the position of each train wheel set in relation to the compensation capacitor and the rail line, and constructed the model of rail-out voltage and current flowing through the first wheel set which is used as the short circuit current based on the four-terminal network and transmission line theory.  Fenxia Tian\cite{tian2020} established the train short circuit current model of each wheel set based on the transmission line theory and analyzed the difference of signal current flowing through different wheel sets when the train passed through the poor shunting area. Bin Hao\cite{bin2022simulation} utilized the on-state change of  timed switches based on the Simulink tool to simulate the dynamic shunting process of multiple train wheel sets and designed a ZPW-2000 multi-section track circuit simulation model.

However, there are still some shortcomings in the above study. The study in \cite{fu2018research} neglects the electromagnetic induction process between the current flowing through wheel sets and the track circuit reader (TCR) antenna\cite{zhao2013modeling}, which makes the simulation results have deviations. The study in \cite{fu2018research, tian2020, bin2022simulation} do not consider the leakage of signal current to the ground in the transmission of steel rails\cite{zhao2013modeling}, which has certain limitation.

In conclusion, the train shunting impedance has not been adequately studied. Therefore, under the consideration of rail current leakage to ground, this paper firstly models, simulates and verifies the train shunting impedance from the formation and operation process of the train based on transmission line theory. Then, the influencing factors of train shunting impedance and their influence law are analyzed. Finally, the simplified model of train shunting impedance is determined by calculating the structural importance of train wheel sets. Experiments show that the method proposed in this paper has the advantages of higher accuracy and simulation precision.

\section{Model of train shunting impedance \\ based on transmission line theory}
\begin{figure*}[!t]
	\centering
	\includegraphics[width=\linewidth]{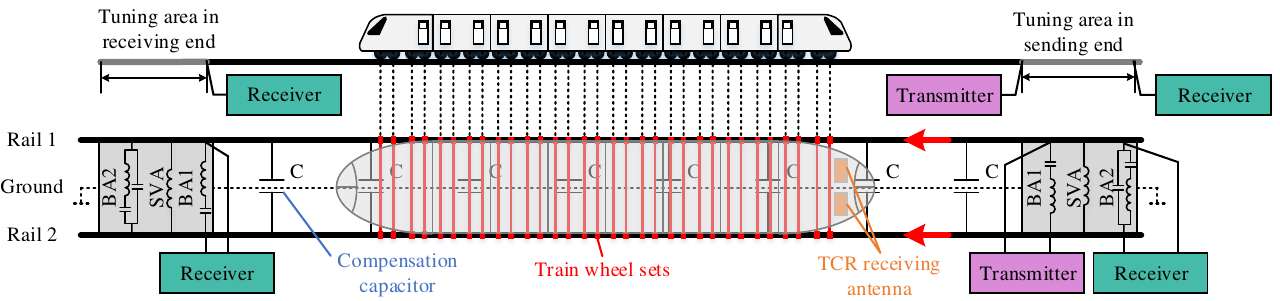}
	\caption{Schematic diagram of the JTC in occupied state.}
	\label{fig_sd_jtc}
\end{figure*}

As shown in Fig. \ref{fig_sd_jtc}, the JTC consists of compensation capacitors, tuning units (BA1 and BA2) and air core coils (SVA) installed between two steel rails, which affect the train shunting impedance. As the train enters and departs JTC, the relative position between train wheel sets and parallel equipment is constantly changing, resulting in the change of train shunting impedance.

In this study, firstly, the equivalent six-terminal network (ESTN) model of rail line, parallel equipment (compensation capacitors, tuning area equipment and train wheel sets) and rail-wheel unit (which consists of rail line and train wheel sets between two adjacent parallel equipment) are established based on the transmission line theory. Then, the tuning area in receiving end, the  main track circuit after the train shunting point, the main track circuit before the train shunting point, and the tuning area in sending end are modeled according to the train operation direction and the train shunting point. Finally, the ESTN model of the train shunting impedance is established based on the above modeling process.

\subsection{Model of rail line}

\begin{figure}[!t]
	\centering
	\includegraphics[width=2in]{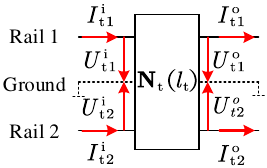}
	\caption{ESTN model of rail line.}
	\label{fig_estn_rail}
\end{figure}

The rail line is equivalent to a circuit composed of three wires: two rails and the ground considering ground leakage of the JTC signal current, whose ESTN is shown in Fig. \ref{fig_estn_rail}.

Here, $\mathbf{N}_\text{t} (l_\text{t})$ is the EFTN of rail line with length $l_\text{t}$. $U^\text{i}_\text{t1}$, $U^\text{i}_\text{t2}$,  $I^\text{i}_\text{t1}$ and $I^\text{i}_\text{t2}$ are the voltage to ground and current on the input terminals, respectively. Correspondingly, $U^\text{o}_\text{t1}$, $U^\text{o}_\text{t2}$,  $I^\text{o}_\text{t1}$ and $I^\text{o}_\text{t2}$ are the voltage to ground and current on the output terminals, respectively. According to the transmission line theory\cite{ch2006electromagnetic}, the ESTN $\mathbf{N}_\text{t} (l_\text{t})$ satisfies
\begin{equation}
\begin{bmatrix}
    U^\text{i}_\text{t1} & U^\text{i}_\text{t2} & I^\text{i}_\text{t1} & I^\text{i}_\text{t2}
\end{bmatrix}^\top = \mathbf{N}_\text{t} (l_\text{t})
\begin{bmatrix}
    U^\text{o}_\text{t1} & U^\text{o}_\text{t2} & I^\text{o}_\text{t1} & I^\text{o}_\text{t2}
\end{bmatrix}^\top
\end{equation}

\begin{figure}[!t]
	\centering
	\includegraphics[width=\linewidth]{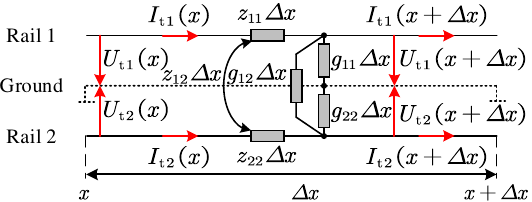}
	\caption{Equivalent circuit of rail line with length $\varDelta x$ at position $x$.}
	\label{fig_ec_rail}
\end{figure}

The equivalent circuit of rail line with length $\varDelta x$ at any position $x$ is shown in Fig. \ref{fig_ec_rail}.

Here, $U_\text{t1}(x)$, $U_\text{t2}(x)$, $I_\text{t1}(x)$ and $I_\text{t2}(x)$ are the voltage to ground and current on the two rails at position $x$, respectively. Correspondingly, $U_\text{t1}(x+\varDelta x)$, $U_\text{t2}(x+\varDelta x)$, $I_\text{t1}(x+\varDelta x)$ and $I_\text{t2}(x+\varDelta x)$ are the voltage to ground and current on the two rails at position $x+\varDelta x$, respectively. $z_{11}$, $z_{12}$, $g_{11}$ and $g_{12}$ are the unit impedance and unit admittance of the two rails, respectively. $z_{12}$ and $g_{12}$ are the unit mutual impedance and unit mutual admittance of the two rails, respectively.

According to Kirchhoff's low\cite{gabelli2006violation}, we have
\begin{equation}
\left\{ \begin{aligned}
	\frac{U_\text{t1}(x+\varDelta x) - U_\text{t1}(x)}{\varDelta x} &= -z_{11}I_\text{t1}(x) - z_{12}I_\text{t2}(x) \\
	\frac{U_\text{t2}(x+\varDelta x) - U_\text{t2}(x)}{\varDelta x} &= -z_{22}I_\text{t2}(x) - z_{12}I_\text{t1}(x) \\
	\frac{I_\text{t1}(x+\varDelta x) - I_\text{t1}(x)}{\varDelta x} &= -g_{11}U_\text{t1}(x+\varDelta x) \\
																	 &-g_{12}(U_\text{t1}(x+\varDelta x) - U_\text{t2}(x+\varDelta x)) \\
	\frac{I_\text{t2}(x+\varDelta x) - I_\text{t2}(x)}{\varDelta x} &= -g_{22}U_\text{t2}(x+\varDelta x) \\
																	 &-g_{12}(U_\text{t2}(x+\varDelta x) - U_\text{t1}(x+\varDelta x)) \\
\end{aligned} \right. 
\end{equation}

Letting $\varDelta x\rightarrow0$, the transmission line equation of rail line can be expressed as
\begin{equation}
\label{equa_dif}
\left\{ \begin{aligned}
	&\begin{bmatrix}
		\frac{\text{d} U_\text{t1}(x)}{\text{d} x} & \frac{\text{d} U_\text{t2}(x)}{\text{d} x}
	\end{bmatrix}^\top = \mathbf{z}_\text{t} \begin{bmatrix}
		I_\text{t1}(x) & I_\text{t2}(x)
	\end{bmatrix}^\top \\
	&\begin{bmatrix}
		\frac{\text{d} I_\text{t1}(x)}{\text{d} x} & \frac{\text{d} I_\text{t2}(x)}{\text{d} x}
	\end{bmatrix}^\top = \mathbf{g}_\text{t} \begin{bmatrix}
		U_\text{t1}(x) & U_\text{t2}(x)
	\end{bmatrix}^\top \\
\end{aligned}
\right.
\end{equation}
\noindent where $\mathbf{z}_\text{t}$ and $\mathbf{g}_\text{t}$ are the impedance matrix and admittance matrix of rail line, respectively. That is
\begin{equation}
\mathbf{z}_\text{t}=\begin{bmatrix}
	-z_{11} & -z_{12} \\
	-z_{12} & -z_{22}
\end{bmatrix} 
\end{equation}
\begin{equation}
	\mathbf{g}_\text{t}=\begin{bmatrix}
		-g_{11}-g_{12} & g_{12} \\
		g_{12} & -g_{22}-g_{12}
	\end{bmatrix} 
\end{equation}

Solving the differential equations shown in (\ref{equa_dif}), the general solution of $U_\text{t1}$ and $U_\text{t2}$ can be expressed as
\begin{equation}
	\label{equa_solve_u}
	\begin{aligned}
		\begin{bmatrix}
			U_\text{t1}(x) \\ U_\text{t2}(x)
		\end{bmatrix} &= \mathbf{H}_\text{t} \left(
		\begin{bmatrix}
			\cosh(\sqrt{\lambda _\text{t1}}x) & 0 \\
			0 & \cosh(\sqrt{\lambda _\text{t2}}x)
		\end{bmatrix}
		\begin{bmatrix}
			a_\text{t1} \\ b_\text{t1}
		\end{bmatrix} \right. \\
		&\left. + \begin{bmatrix}
			\sinh(\sqrt{\lambda _\text{t1}}x) & 0 \\
			0 & \sinh(\sqrt{\lambda _\text{t2}}x)
		\end{bmatrix}
		\begin{bmatrix}
			a_\text{t2} \\ b_\text{t2}
		\end{bmatrix}
		\right)
	\end{aligned}
\end{equation}
\noindent where $a_\text{t1}$, $b_\text{t1}$, $a_\text{t2}$ and $b_\text{t2}$ are corresponding indeterminate coefficients. $\lambda_\text{t1}$ and $\lambda_\text{t2}$ are the eigenvalues of $\mathbf{z}_\text{t}\mathbf{g}_\text{t}$. $\mathbf{H}_\text{t}$ is the matrix composed of the corresponding eigenvectors. Substituting (\ref{equa_solve_u}) into (\ref{equa_dif}), it arrives at
\begin{equation}
	\label{equa_solve_i}
	\begin{aligned}
		\begin{bmatrix}
			I_\text{t1}(x) \\ I_\text{t2}(x)
		\end{bmatrix} &= \mathbf{z}_\text{t}^{-1} \mathbf{H}_\text{t} \\ 
		&\left( \begin{bmatrix}
			\sqrt{\lambda_{t1}}\sinh(\sqrt{\lambda_{t1}}x) & 0 \\
			0 & \sqrt{\lambda_{t2}}\sinh(\sqrt{\lambda_{t2}}x)
		\end{bmatrix}
		\begin{bmatrix}
			a_\text{t1} \\ b_\text{t1}
		\end{bmatrix} \right. \\
		&\left. + \begin{bmatrix}
			\sqrt{\lambda_{t1}}\cosh(\sqrt{\lambda_{t1}}x) & 0 \\
			0 & \sqrt{\lambda_{t2}}\cosh(\sqrt{\lambda_{t2}}x)
		\end{bmatrix}
		\begin{bmatrix}
			a_\text{t2} \\ b_\text{t2}
		\end{bmatrix}
		\right)
	\end{aligned}
\end{equation}

According to the boundary conditions of JTC, the ESTN $\mathbf{N}_\text{t} (l_\text{t})$ can be given by 
\begin{equation}
\label{equa_estn_rail}
\begin{aligned}
	&\mathbf{H}_\text{t}(l_\text{t})= \\
	&\begin{bmatrix}
		\mathbf{H}_\text{t}\mathbf{D}_\text{t1}(l_\text{t})\mathbf{H}_\text{t}^{-1} & \mathbf{H}_\text{t}\mathbf{D}_\text{t2}(l_\text{t})\mathbf{D}_\text{t3}^{-1}\mathbf{H}_\text{t}^{-1}\mathbf{z}_\text{t} \\
		\mathbf{z}_\text{t}^{-1}\mathbf{H}_\text{t}\mathbf{D}_\text{t3}\mathbf{D}_\text{t2}(l_\text{t})\mathbf{H}_\text{t}^{-1} & 
		\mathbf{z}_\text{t}^{-1}\mathbf{H}_\text{t}\mathbf{D}_\text{t3}\mathbf{D}_\text{t1}(l_\text{t})\mathbf{D}_\text{t3}^{-1}\mathbf{H}_\text{t}^{-1}\mathbf{z}_\text{t}
	\end{bmatrix}^{-1}
\end{aligned}
\end{equation}
\noindent where $\mathbf{D}_\text{t1}(l_\text{t})$, $\mathbf{D}_\text{t2}(l_\text{t})$ and $\mathbf{D}_\text{t3}$ can be expressed as 
\begin{equation}
	\mathbf{D}_\text{t1}(l_\text{t})=\begin{bmatrix}
		\cosh(\sqrt{\lambda_\text{t1}}l_\text{t}) & 0 \\
		0 & \cosh(\sqrt{\lambda_\text{t2}}l_\text{t})
	\end{bmatrix}
\end{equation}
\begin{equation}
	\mathbf{D}_\text{t2}(l_\text{t})=\begin{bmatrix}
		\sinh(\sqrt{\lambda_\text{t1}}l_\text{t}) & 0 \\
		0 & \sinh(\sqrt{\lambda_\text{t2}}l_\text{t})
	\end{bmatrix}
\end{equation}
\begin{equation}
	\mathbf{D}_\text{t3}=\begin{bmatrix}
		\sqrt{\lambda_\text{t1}} & 0 \\
		0 & \sqrt{\lambda_\text{t2}}
	\end{bmatrix}
\end{equation}
\subsection{Model of parallel equipment}
\begin{figure}[!t]
	\centering
	\includegraphics[width=2in]{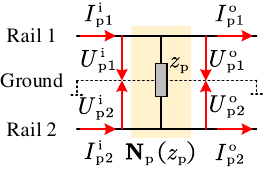}
	\caption{ESTN model of parallel equipment.}
	\label{fig_estm_p}
\end{figure}

The parallel equipment between the steel rails consists of compensation capacitors and tuning area equipment (BA1, BA2 and SVA). Setting the equivalent impedance of the parallel equipment is $z_\text{p}$ without loss of generality, the ESTN of parallel equipment $\mathbf{N}_\text{p}(z_\text{p})$ is shown in Fig. \ref{fig_estm_p}.

Here, $U_\text{p1}^\text{i}$, $U_\text{p2}^\text{i}$, $I_\text{p1}^\text{i}$ and $I_\text{p2}^\text{i}$ are the voltage to ground and current on the input terminals, respectively. Correspondingly, $U_\text{p1}^\text{o}$, $U_\text{p2}^\text{o}$, $I_\text{p1}^\text{o}$ and $I_\text{p2}^\text{o}$ are the voltage to ground and current on the output terminals, respectively. According to the transmission line theory, the ESTN $\mathbf{N}_\text{p}(z_\text{p})$ satisfies
\begin{equation}
	\label{equa_trans_p}
	\begin{bmatrix}
		U_\text{p1}^\text{i} & U_\text{p2}^\text{i} & I_\text{p1}^\text{i} & I_\text{p2}^\text{i}
	\end{bmatrix}^\top = \mathbf{N}_\text{p}(z_\text{p})
	\begin{bmatrix}
		U_\text{p1}^\text{o} & U_\text{p2}^\text{o} & I_\text{p1}^\text{o} & I_\text{p2}^\text{o}
	\end{bmatrix}^\top
\end{equation}

According to Kirchhoff's low\cite{gabelli2006violation}, we have
\begin{equation}
	\label{equa_krich_p}
	\left\{
	\begin{aligned}
		& U_\text{p1}^\text{i}=U_\text{p1}^\text{o}, U_\text{p2}^\text{i}=U_\text{p2}^\text{o} \\
		& I_\text{p1}^\text{i}-I_\text{p1}^\text{o} = I_\text{p2}^\text{i}-I_\text{p2}^\text{o} \\
		& \frac{U_\text{p1}^\text{i} - U_\text{p2}^\text{i}}{z_\text{p}} = \frac{I_\text{p1}^\text{i} - I_\text{p1}^\text{o}}{2} + \frac{I_\text{p2}^\text{i} - I_\text{p2}^\text{o}}{2}
	\end{aligned}
	\right.
\end{equation}

Using (\ref{equa_trans_p}) and (\ref{equa_krich_p}), the ESTN $\mathbf{N}_\text{p}(z_\text{p})$ can be expressed as
\begin{equation}
    \label{equa_estn_p}
	\mathbf{N}_\text{p}(z_\text{p}) = \begin{bmatrix}
		\mathbf{E}_{2\times2} & \mathbf{O}_{2\times2} \\
		\begin{matrix}
		1/z_\text{p} & -1/z_\text{p} \\
		-1/z_\text{p} & 1/z_\text{p}
		\end{matrix} & \mathbf{E}_{2\times2}
	\end{bmatrix}
\end{equation}
\noindent where $\mathbf{E}_{2\times2}$ and $\mathbf{O}_{2\times2}$ are the unit matrix of $2\times 2$ and the zero matrix of $2\times 2$, respectively.
\subsection{Model of rail-wheel unit}
\begin{figure}[!t]
	\centering
	\includegraphics[width=3in]{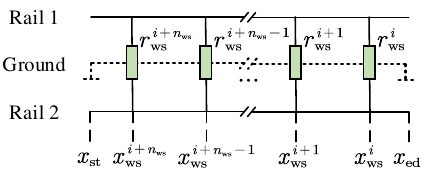}
	\caption{Structure of rail-wheel unit.}
	\label{fig_struc_rwunit}
\end{figure}

In order to comprehensively consider the situation when train occupies or doesn't occupy JTC, the section of rail line between two adjacent parallel equipment is modeled in the form of rail-wheel unit shown in Fig. \ref{fig_struc_rwunit}. 

Here, take CRH380B series high-speed electric multiple units (EMUs) shown in Fig. \ref{fig_sd_jtc} as an example, which have 32 wheel sets. Let $\mathbf{x}_\text{rw}$ and $\mathbf{r}_\text{rw}$ denote the position vector and impedance vector along the transmission direction of JTC signal, respectively. That is
\begin{equation}
\begin{aligned}
	&\mathbf{x}_\text{rw}=[x_\text{st}, x_\text{ws}^{i+n_\text{ws}}, x_\text{ws}^{i+n_\text{ws}-1}, \ldots, x_\text{ws}^i, x_\text{ed}]^\top \\
	&\overset{\text{denoted as}}{=} [x_\text{ws,1},x_\text{ws,2},\ldots,x_{\text{ws,}n_\text{ws}+2}]^\top
\end{aligned}
\end{equation}
\begin{equation}
	\begin{aligned}
		&\mathbf{r}_\text{rw}=[r_\text{ws}^{i+n_\text{ws}}, r_\text{ws}^{i+n_\text{ws}-1}, \ldots, r_\text{ws}^i]^\top \\
		&\overset{\text{denoted as}}{=} [r_\text{ws,1},r_\text{ws,2},\ldots,r_{\text{ws,}n_\text{ws}}]^\top
	\end{aligned}
\end{equation}
\noindent where $n_\text{ws}$ is the number of train wheel sets in rail-wheel unit. $x_\text{ws}^i$ and $r_\text{ws}^i$ are the position and impedance of i-th wheel set from the head vehicle, respectively. $x_\text{st}$ and $x_\text{ed}$ are the start and end position of rail-wheel unit, respectively. Using (\ref{equa_estn_rail}) and (\ref{equa_estn_p}), the ESTN of rail-wheel unit $\mathbf{N}_\text{rw}(\mathbf{x}_\text{rw}, \mathbf{r}_\text{rw})$ can be given by
\begin{equation}
	\label{equa_estn_rw}
	\begin{aligned}
		\mathbf{N}_\text{rw}(\mathbf{x}_\text{rw}, \mathbf{r}_\text{rw}) &= \mathbf{N}_\text{t}(x_\text{ws,2}-x_\text{ws,1}) \\
		& \times \prod_{j=2}^{n_\text{ws}-1} \mathbf{N}_\text{p}(r_{\text{ws,}j-1}) \mathbf{N}_\text{t}(x_{\text{ws,}j+1} - x_{\text{ws,}j})
	\end{aligned}
\end{equation}
\subsection{Model of train shunting impedance}
\begin{figure*}[!t]
	\centering
	\includegraphics[width=\linewidth]{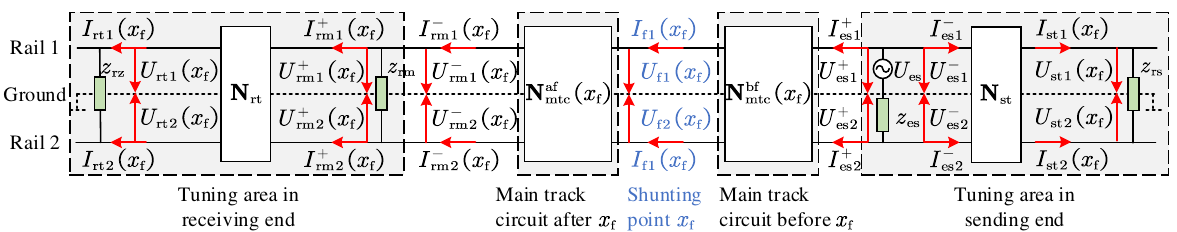}
	\caption{ESTN model of JTC in occupied state.}
	\label{fig_estn_jtc}
\end{figure*}

Based on the operation process of the train, the calculation situation of train shunting impedance can be mainly divided into two cases: fully and partially occupancy of JTC by the train. Due to the limited space, this paper takes the situation when the train fully occupies the main track as an example. Under this situation, the corresponding ESTN model of JTC in occupied state is shown in Fig. \ref{fig_estn_jtc}.

Setting the shunting point is $x_\text{f}$, each component of JTC can be modeled. The specific modeling steps are as follows:
\subsubsection{Model of tuning area in receiving end}
Firstly, as for the tuning unit BA2, according th Kirchhoff's low\cite{gabelli2006violation}, the equivalent impedance of BA2 $z_\text{rz}$ satisfies
\begin{equation}
\begin{aligned}
    &\begin{bmatrix}
        1 & -1 & -z_\text{rz}/2 & z_\text{rz}/2 \\
        0 & 0 & 1 & 1
    \end{bmatrix} \\
    &\times \begin{bmatrix}
        U_\text{rt1}(x_\text{f}) & U_\text{rt2}(x_\text{f}) & I_\text{rt1}(x_\text{f}) & I_\text{rt2}(x_\text{f})
    \end{bmatrix}^\top=\mathbf{O}_{2\times 1}
\end{aligned}
\end{equation}
\noindent where $U_\text{rt1}$, $U_\text{rt2}$, $I_\text{rt1}$ and $I_\text{rt2}$ are the voltage to ground and current on steel rails at both ends of BA2.

Secondly, letting $\mathbf{N}_\text{rt}$ denotes the ESTN of the tuning area in receiving end expect for BA1 and BA2, the ESTN $\mathbf{N}_\text{rt}$ can be given by
\begin{equation}
    \mathbf{N}_\text{rt} = \mathbf{N}_\text{t}(l_\text{st}/2)\mathbf{N}_\text{p}(z_\text{SVA})\mathbf{N}_\text{t}(l_\text{st}/2)
\end{equation}
\noindent where $z_\text{SVA}$ is the equivalent impedance of the air core coil SVA. $l_\text{st}$ is the length of tuning area. According to the transmission line theory\cite{ch2006electromagnetic}, the ESTN $\mathbf{N}_\text{rt}$ satisfies
\begin{equation}
\begin{aligned}
    &\begin{bmatrix}
        U_\text{rm1}^+(x_\text{f}) & U_\text{rm2}^+(x_\text{f}) & I_\text{rm1}^+(x_\text{f}) & I_\text{rm2}^+(x_\text{f})
    \end{bmatrix}^\top \\
    &= \mathbf{N}_\text{rt} \begin{bmatrix}
        U_\text{rt1}(x_\text{f}) & U_\text{rt2}(x_\text{f}) & I_\text{rt1}(x_\text{f}) & I_\text{rt2}(x_\text{f})
    \end{bmatrix}^\top
\end{aligned}
\end{equation}

Finally, considering that the tuning unit BA1 is connected to the main track circuit receiving channel, BA1 is equated to parallel equipment with impedance $z_\text{rm}$, which denotes the input impedance from BA1 to the receiver. Using (\ref{equa_estn_p}), the ESTN of BA1 can be expressed as $\mathbf{N}_\text{p}(z_\text{rm})$. According to the transmission line theory\cite{ch2006electromagnetic}, the ESTN $\mathbf{N}_\text{p}(z_\text{rm})$ satisfies
\begin{equation}
\begin{aligned}
    &\begin{bmatrix}
        U_\text{rm1}^-(x_\text{f}) & U_\text{rm2}^-(x_\text{f}) & I_\text{rm1}^-(x_\text{f}) & I_\text{rm2}^-(x_\text{f})
    \end{bmatrix}^\top \\
    &= \mathbf{N}_\text{p}(z_\text{rm}) \begin{bmatrix}
        U_\text{rm1}^+(x_\text{f}) & U_\text{rm2}^+(x_\text{f}) & I_\text{rm1}^+(x_\text{f}) & I_\text{rm2}^+(x_\text{f})
    \end{bmatrix}^\top
\end{aligned}
\end{equation}
\subsubsection{Model of main track circuit after $x_\text{f}$}
\begin{figure}[!t]
	\centering
	\includegraphics[width=\linewidth]{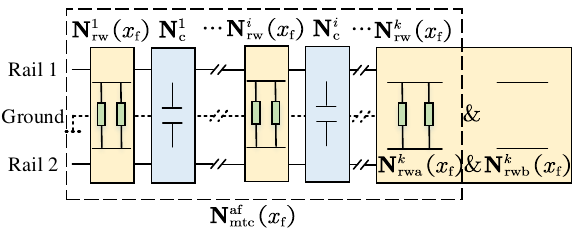}
	\caption{ESTN model of the main track circuit after $x_\text{f}$.}
	\label{fig_estn_mtc_a}
\end{figure}

The ESTN model of the main track circuit after shunting point $x_\text{f}$ is shown in Fig. \ref{fig_estn_mtc_a}, which is composed of several rail-wheel units and compensation capacitors.

Here, $\mathbf{N}_\text{mtc}^\text{af}(x_\text{f})$ is the ESTN of the main track circuit after shunting point $x_\text{f}$. $\mathbf{N}_\text{rw}^i$ and $\mathbf{N}_\text{c}^i$ are the ESTN of i-th rail-wheel unit and compensation capacitor from the receiving end, respectively. 

Assuming that shunting point $x_\text{f}$ is in the $k$-th rail-wheel unit, the ESTN $\mathbf{N}_\text{rw}^k$ can be divided into two parts: $\mathbf{N}_\text{rwa}^k$ and $\mathbf{N}_\text{rwb}^k$. $\mathbf{N}_\text{rwb}^k$ is the ESTN of rail line from $x_\text{ed}$ to $x_\text{f}$ (without any train wheel sets). $\mathbf{N}_\text{rwa}^k$ is the rest part of this rail-wheel unit.

Using (\ref{equa_estn_p}) and (\ref{equa_estn_rw}), the ESTN $\mathbf{N}_\text{mtc}^\text{af}(x_\text{f})$ can be given by
\begin{equation}
	\mathbf{N}_\text{mtc}^\text{af}(x_\text{f}) = \mathbf{N}_\text{rwa}^k(x_\text{f})\times \prod_{j=k-1}^{1}\mathbf{N}_\text{c}^j \mathbf{N}_\text{rw}^j(x_\text{f})
\end{equation}

According to the transmission line theory\cite{ch2006electromagnetic}, the ESTN $\mathbf{N}_\text{mtc}^\text{af}(x_\text{f})$ satisfies
\begin{equation}
	\begin{aligned}
		&\begin{bmatrix}
			U_\text{f1}(x_\text{f}) & U_\text{f2}(x_\text{f}) & I_\text{f1}(x_\text{f}) & I_\text{f2}(x_\text{f})
		\end{bmatrix}^\top \\
		&= \mathbf{N}_\text{mtc}^\text{af}(x_\text{f}) \begin{bmatrix}
			U_\text{rm1}^-(x_\text{f}) & U_\text{rm2}^-(x_\text{f}) & I_\text{rm1}^-(x_\text{f}) & I_\text{rm2}^-(x_\text{f})
		\end{bmatrix}^\top
	\end{aligned}
\end{equation}
\noindent where $U_\text{f1}(x_\text{f})$, $U_\text{f2}(x_\text{f})$, $I_\text{f1}(x_\text{f})$ and $I_\text{f2}(x_\text{f})$ are the voltage to ground and current on steel rails under the first train wheel set.
\subsubsection{Model of main track circuit before $x_\text{f}$}
\begin{figure}[!t]
	\centering
	\includegraphics[width=\linewidth]{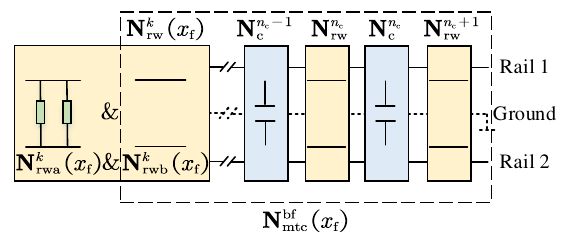}
	\caption{ESTN model of the main track circuit before $x_\text{f}$.}
	\label{fig_estn_mtc_b}
\end{figure}

The ESTN model of the main track circuit before shunting point $x_\text{f}$ is shown in Fig. \ref{fig_estn_mtc_b}, which is composed of several sections of rail line and compensation capacitors.

Here, $\mathbf{N}_\text{mtc}^\text{bf}(x_\text{f})$ is the ESTN of the main track circuit before shunting point $x_\text{f}$. $n_\text{c}$ is the number of compensation capacitors.

Using (\ref{equa_estn_p}) and (\ref{equa_estn_rw}), the ESTN $\mathbf{N}_\text{mtc}^\text{bf}(x_\text{f})$ can be given by
\begin{equation}
	\mathbf{N}_\text{mtc}^\text{bf}(x_\text{f}) = \prod_{j=n_\text{c}+1}^{k+1}\mathbf{N}_\text{rw}^j \mathbf{N}_\text{c}^{j-1} \times \mathbf{N}_\text{rwb}^k(x_\text{f})
\end{equation}
\subsubsection{Model of tuning area in sending end}
Firstly, as for the tuning unit BA1, according to Thevenin's theorem\cite{brittain1990thevenin}, the sending channel can be equated to a voltage source $U_\text{es}$ and impedance $z_\text{es}$ in series\cite{zhao2013modeling}. According to Kirchhoff's low\cite{gabelli2006violation} and transmission line theory\cite{ch2006electromagnetic}, we have
\begin{equation}
\begin{aligned}
	&\begin{bmatrix}
		\mathbf{E}_{2\times2} & \mathbf{O}_{2\times2} \\
		\begin{matrix}
			0 & 0 \\
			1 & -1
		\end{matrix} & \begin{matrix}
			1 & 1 \\
			z_\text{es}/2 & -z_\text{es}/2
		\end{matrix}
	\end{bmatrix} \times 
	\begin{bmatrix}
		U_\text{es1}^+ \\ U_\text{es2}^+ \\ I_\text{es1}^+ \\ I_\text{es2}^+
	\end{bmatrix} \\
	&+\begin{bmatrix}
		-\mathbf{E}_{2\times2} & \mathbf{O}_{2\times2} \\
		\mathbf{O}_{2\times 2} & 
		\begin{matrix}
			1 & 1 \\
			z_\text{es}/2 & -z_\text{es}/2
		\end{matrix}
	\end{bmatrix} \times 
	\begin{bmatrix}
		U_\text{es1}^- \\ U_\text{es2}^- \\ I_\text{es1}^- \\ I_\text{es2}^-
	\end{bmatrix} = 
	\begin{bmatrix}
		0 \\ 0 \\ 0 \\ U_\text{es}
	\end{bmatrix}
\end{aligned}
\end{equation}
\begin{equation}
	\begin{aligned}
		&\begin{bmatrix}
			U_\text{es1}^+ & U_\text{es2}^+ & I_\text{es1}^+ & I_\text{es2}^+
		\end{bmatrix}^\top \\
		&= \mathbf{N}_\text{mtc}^\text{bf}(x_\text{f}) \begin{bmatrix}
			U_\text{f1}(x_\text{f}) & U_\text{f2}(x_\text{f}) & I_\text{f1}(x_\text{f}) & I_\text{f2}(x_\text{f})	
		\end{bmatrix}^\top
	\end{aligned}
\end{equation}

Secondly, letting $\mathbf{N}_\text{st}$ denotes the ESTN of the tuning area in sending end expect for BA1 and BA2, the ESTN $\mathbf{N}_\text{st}$ can be given by 
\begin{equation}
	\mathbf{N}_\text{st} = \mathbf{N}_\text{t}(l_\text{st}/2)\mathbf{N}_\text{p}(z_\text{SVA})\mathbf{N}_\text{t}(l_\text{st}/2)
\end{equation}

According to the transmission line theory\cite{ch2006electromagnetic}, the ESTN $\mathbf{N}_\text{st}$ satisfies
\begin{equation}
	\begin{aligned}
		&\begin{bmatrix}
			U_\text{es1}^- & U_\text{es2}^- & I_\text{es1}^- & I_\text{es2}^-
		\end{bmatrix}^\top \\
		&= \mathbf{N}_\text{st} \begin{bmatrix}
			U_\text{st1}(x_\text{f}) & U_\text{st2}(x_\text{f}) & I_\text{st1}(x_\text{f}) & I_\text{st2}(x_\text{f})
		\end{bmatrix}^\top
	\end{aligned}
\end{equation}

Finally, considering that the tuning unit BA2 is connected to the short track circuit receiving channel, BA2 is equated to parallel equipment with impedance $z_\text{rs}$, which denotes the input impedance from BA2 to the receiver. According th Kirchhoff's low\cite{gabelli2006violation}, the equivalent impedance of BA2 $z_\text{rs}$ satisfies
\begin{equation}
	\begin{aligned}
		&\begin{bmatrix}
			1 & -1 & -z_\text{rs}/2 & z_\text{rs}/2 \\
			0 & 0 & 1 & 1
		\end{bmatrix} \\
		&\times \begin{bmatrix}
			U_\text{rs1}(x_\text{f}) & U_\text{rs2}(x_\text{f}) & I_\text{rs1}(x_\text{f}) & I_\text{rs2}(x_\text{f})
		\end{bmatrix}^\top=\mathbf{O}_{2\times 1}
	\end{aligned}
\end{equation}
\subsubsection{Model of train shunting impedance}
On the basis of the above modeling process, $U_\text{f1}(x_\text{f})$, $U_\text{f2}(x_\text{f})$, $I_\text{f1}(x_\text{f})$ and $I_\text{f2}(x_\text{f})$ can be given by
\begin{equation}
\label{equa_ui_fl}
\begin{aligned}
	&\begin{bmatrix}
		U_\text{f1}(x_\text{f}) & U_\text{f2}(x_\text{f}) & I_\text{f1}(x_\text{f}) & I_\text{f2}(x_\text{f})
	\end{bmatrix}^\top= \\
	&\begin{bmatrix}
		\mathbf{N}_4\mathbf{N}_\text{st}^{-1}\mathbf{N}_3^{-1}\mathbf{N}_2\mathbf{N}_\text{mtc}^\text{bf}(x_\text{f}) \\
		\mathbf{N}_1\mathbf{N}_\text{rt}^{-1}(\mathbf{N}_\text{p}(z_\text{rm}))^{-1}(\mathbf{N}_\text{mtc}^\text{af}(x_\text{f}))^{-1}
	\end{bmatrix}^\top 
	\begin{bmatrix}
		\mathbf{N}_4\mathbf{N}_\text{st}^{-1}\mathbf{N}_3^{-1}\mathbf{U}_\text{es} \\
		\mathbf{O}_{2\times1}
	\end{bmatrix}
\end{aligned}
\end{equation}
\noindent where $\mathbf{N}_1$, $\mathbf{N}_2$, $\mathbf{N}_3$, $\mathbf{N}_4$ and $\mathbf{U}_{es}$ can be expressed as
\begin{equation}
	\mathbf{N}_1 = \begin{bmatrix}
		1 & -1 & -z_\text{rz}/2 & z_\text{rz}/2 \\
		0 & 0 & 1 & 1
	\end{bmatrix}
\end{equation}
\begin{equation}
	\mathbf{N}_2 = \begin{bmatrix}
		\mathbf{E}_{2\times2} & \mathbf{O}_{2\times2} \\
		\begin{matrix}
			0 & 0 \\
			1 & -1
		\end{matrix} & \begin{matrix}
			1 & 1 \\
			z_\text{es}/2 & -z_\text{es}/2
		\end{matrix}
	\end{bmatrix}
\end{equation}
\begin{equation}
	\mathbf{N}_3 = \begin{bmatrix}
		-\mathbf{E}_{2\times2} & \mathbf{O}_{2\times2} \\
		\mathbf{O}_{2\times 2} & 
		\begin{matrix}
			1 & 1 \\
			z_\text{es}/2 & -z_\text{es}/2
		\end{matrix}
	\end{bmatrix}
\end{equation}
\begin{equation}
	\mathbf{N}_4 = \begin{bmatrix}
		1 & -1 & -z_\text{rs}/2 & z_\text{rs}/2 \\
		0 & 0 & 1 & 1
	\end{bmatrix}
\end{equation}
\begin{equation}
	\mathbf{U}_{es} = \begin{bmatrix}
		0 & 0 & 0 & U_{es}
	\end{bmatrix}^\top
\end{equation}

Using (\ref{equa_ui_fl}), the train shunting impedance $z_\text{f}(x_\text{f})$ can be given by
\begin{equation}
	z_\text{f}(x_\text{f})=\frac{U_\text{f1}(x_\text{f}) - U_\text{f2}(x_\text{f})}{(I_\text{f1}(x_\text{f}) - I_\text{f2}(x_\text{f}))/2}
\end{equation}
\subsection{Verification and analysis of the model $z_\text{f}(x_\text{f})$}
It is difficult to obtain the resistance of each wheel set and the train shunting impedance in real time during the train operation process. Therefore, the induced voltage signal between TCR and the rail line is utilized to achieve the indirect verification of the model $z_\text{f}(x_\text{f})$ based on the working principle of TCR.

First of all, based on the model constructed above, the simulation of train shunting impedance $z_\text{f}(x_\text{f})$ at each position of JTC is carried out, whose main simulation conditions are as follows: the length of JTC is 789m; carrier frequency of JTC signal is 2300Hz; number of compensation capacitors is 9; normal compensation capacitor value is 46\textmu F; ballast resistance is 6$\Omega\cdot$km; resistance of each wheel is 0.15$\Omega$. 

\begin{figure}[!t]
\centering
\subfloat[]{\includegraphics[width=\linewidth]{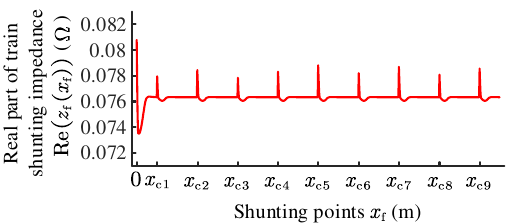}%
	\label{fig_simu_zf_real}}
\hfil
\subfloat[]{\includegraphics[width=\linewidth]{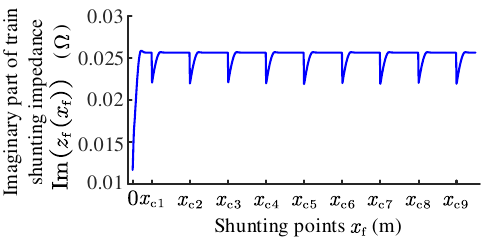}%
	\label{fig_simu_zf_imag}}
\caption{Simulation result of $z_\text{f}(x_\text{f})$. (a) The real part of $z_\text{f}(x_\text{f})$ (b) The imaginary part of $z_\text{f}(x_\text{f})$.}
\label{fig_simu_zf}
\end{figure}

According to the simulation result, $z_\text{f}(x_\text{f})$ is complex impedance, whose real part $\text{Re}(z_\text{f}(x_\text{f}))$ and imaginary part $\text{Im}(z_\text{f}(x_\text{f}))$ are shown in Fig. \ref{fig_simu_zf}.

From Fig. \ref{fig_simu_zf}, it can be seen that the variation of train shunting impedance $z_\text{f}(x_\text{f})$ has observable patterns of regularity, summarized as follows:
\begin{enumerate}
\item{Non-pure resistance: $z_\text{f}(x_\text{f})$ is not a pure resistance, but a complex impedance whose real and imaginary parts are both greater that zero, which can be expressed as a resistance and an inductance in series.}
\item{Local mutation: in the case where the first wheel set coincides the position of compensation capacitors, real part $\text{Re}(z_\text{f}(x_\text{f}))$ and imaginary part $\text{Im}(z_\text{f}(x_\text{f}))$ will show a sudden upward and downward change, respectively. }
\item{General smoothness: after removing the localized mutation, the changes in the rest of real part $\text{Re}(z_\text{f}(x_\text{f}))$ and imaginary part $\text{Im}(z_\text{f}(x_\text{f}))$ tend to be stable, and their overall approximation is a straight line.}
\end{enumerate}

Here, letting $a_1$ denotes the amplitude gain constant during TCR antenna and rail line, $a_2$ denotes the amplitude gain constant of TCR transmission cable\cite{oukhellou2010fault}, the amplitude of TCR induced voltage signal based on train shunting impedance $A_\text{tcr}^\text{zf}(x_\text{f})$ can be given by
\begin{equation}
	A_\text{tcr}^\text{zf}(x_\text{f}) = \left|
		a_1 a_2\frac{U_\text{f1}(x_\text{f})-U_\text{f2}(x_\text{f})}{z_\text{f}(x_\text{f})}
	\right|
\end{equation}

\begin{figure}[!t]
	\centering
	\includegraphics[width=\linewidth]{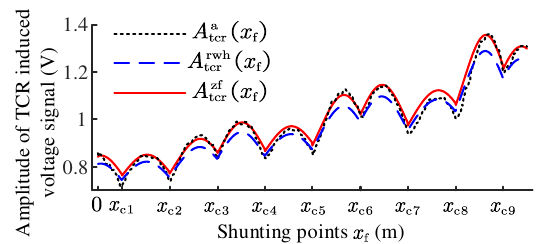}
	\caption{Comparison among $A_\text{tcr}^\text{zf}(x_\text{f})$, $A_\text{tcr}^\text{rwh}(x_\text{f})$ and $A_\text{tcr}^\text{a}(x_\text{f})$}
	\label{fig_compare_signal}
\end{figure}

According to \cite{zhao2013modeling}, the amplitude of TCR induced voltage signal based on resistance of the first wheel set $A_\text{tcr}^\text{rwh}(x_\text{f})$ can be calculated. The results of comparison among $A_\text{tcr}^\text{zf}(x_\text{f})$, $A_\text{tcr}^\text{rwh}(x_\text{f})$ and the actual signal $A_\text{tcr}^\text{a}(x_\text{f})$ is shown in Fig. \ref{fig_compare_signal}.

Taking $A_\text{tcr}^\text{a}(x_\text{f})$ as a standard, the goodness-of-fit of $A_\text{tcr}^\text{zwh}(x_\text{f})$ is as follows: $\text{SSE}$=0.8441, $\text{R-square}$=0.9547, $\text{RMSE}$=0.0333. The goodness-of-fit of $A_\text{tcr}^\text{rf}(x_\text{f})$ is as follows: $\text{SSE}$=0.5798, $\text{R-square}$=0.9689, $\text{RMSE}$=0.0726. In contrast, using train shunting impedance $z_\text{f}(x_\text{f})$ can make the amplitude of TCR induced voltage signal closer to the actual signal, which indirectly indicates the accuracy of the model of train shunting impedance in this study.
\section{Analysis of Influence on train shunting impedance}
In this section, the influence low of train wheel set resistance, compensation capacitor, ballast resistance and rail line impedance on train shunting impedance $z_\text{f}(x_\text{f})$ are analyzed.
\subsection{Influence of train wheel set resistance}
\begin{figure}[!t]
	\centering
	\subfloat[]{\includegraphics[width=3.0in]{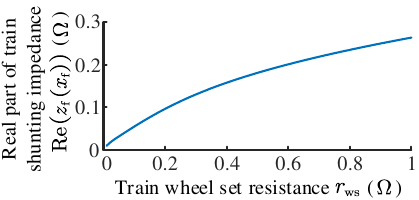}%
		\label{fig_influ_rf_real}}
	\hfil
	\subfloat[]{\includegraphics[width=3.0in]{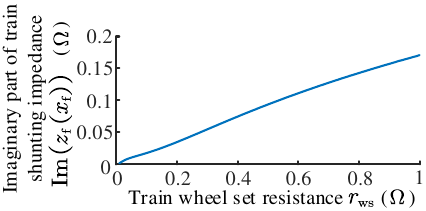}%
		\label{fig_influ_rf_imag}}
	\caption{Relationship between train shunting impedance $z_\text{f}(x_\text{f})$ and train wheel set resistance $r_\text{ws}$. (a) The real part of $z_\text{f}(x_\text{f})$ (b) The imaginary part of $z_\text{f}(x_\text{f})$.}
	\label{fig_influ_rf}
\end{figure}

Based on the simulation condition of Fig. \ref{fig_simu_zf}, the resistance of each wheel set is set to the same value $r_\text{ws}$ with a variation range of $[0.01\Omega,1\Omega]$ and a variation step of $0.01\Omega$. The overall steady state value of real part $\text{Re}(z_\text{f}(x_\text{f}))$ and imaginary part $\text{Im}(z_\text{f}(x_\text{f}))$ are shown in Fig. \ref{fig_influ_rf_real}.
Regression analysis of Fig. \ref{fig_influ_rf_real} yields the corresponding regression model, that is
\begin{equation}
	\label{equa_reg_model_real}
	\text{Re}(z_\text{f}(x_\text{f}))=1/(a_\text{ws}+b_\text{ws}/r_\text{ws})
\end{equation}
\begin{equation}
	\label{equa_reg_model_imag}
	\text{Im}(z_\text{f}(x_\text{f}))=1/(c_\text{ws}+d_\text{ws}/r_\text{ws})
\end{equation}
\noindent where $a_\text{ws}$=2.167, $b_\text{ws}$=1.665, $c_\text{ws}$=0.661 and $d_\text{ws}$=5.095 are the parameters of regression model. The corresponding goodness-of-fit of (\ref{equa_reg_model_real}) is as follows: $\text{SSE}$=$2.0396\times10^{-4}$, $\text{R-square}$=0.9996, $\text{RMSE}$=0.0014. And the corresponding goodness-of-fit of (\ref{equa_reg_model_imag}) is as follows: $\text{SSE}$=$3.3942\times10^{-4}$, $\text{R-square}$=0.9987, $\text{RMSE}$=0.0019. From the goodness-of-fit, it can be seen that there is nonlinear monotonically increasing relationship between train wheel set resistance and train shunting impedance.
\subsection{Influence of compensation capacitor}
\begin{figure}[!t]
	\centering
	\subfloat[]{\includegraphics[width=\linewidth]{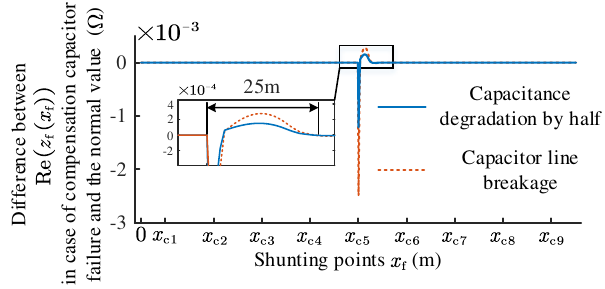}%
		\label{fig_influ_cap_real}}
	\hfil
	\subfloat[]{\includegraphics[width=\linewidth]{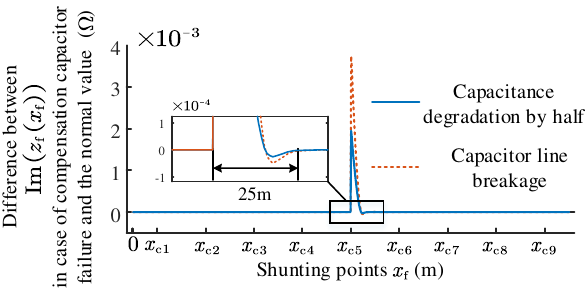}%
		\label{fig_influ_cap_imag}}
	\caption{Relationship between train shunting impedance $z_\text{f}(x_\text{f})$ and the compensation capacitor. (a) The real part of $z_\text{f}(x_\text{f})$ (b) The imaginary part of $z_\text{f}(x_\text{f})$.}
	\label{fig_influ_cap}
\end{figure}

Based on the simulation condition of Fig. \ref{fig_simu_zf}, set up the faults such as capacitor line breakage and capacitance degradation by half are set up respectively, calculate the corresponding train shunting impedance $z_\text{f}(x_\text{f})$, and subtract the real part $\text{Re}(z_\text{f}(x_\text{f}))$ and the imaginary part $\text{Im}(z_\text{f}(x_\text{f}))$ from the normal one shown in Fig. \ref{fig_simu_zf}. The results are shown in Fig. \ref{fig_influ_cap}.

From Fig. \ref{fig_influ_cap}, the influence of compensation capacitor on $z_\text{f}(x_\text{f})$ is bounded, and its influence is only limited to the position of compensation capacitors and its front, i.e., in the direction of the JTC sending end, with in a range of about 25m. When the train shunting point $x_\text{f}$ coincides with the position of compensation capacitors, $z_\text{f}(x_\text{f})$ undergoes a sudden change, which is manifested by the upward and downward sudden change pulses in its real part $\text{Re}(z_\text{f}(x_\text{f}))$ and imaginary part $\text{Im}(z_\text{f}(x_\text{f}))$, and the intensity of the sudden change is proportional to the capacitance of compensation capacitors. Then, as the train operates, the influence of compensation capacitor decreases rapidly, causing the values of $\text{Re}(z_\text{f}(x_\text{f}))$ and $\text{Im}(z_\text{f}(x_\text{f}))$ to decay and increase rapidly, and reaching an overall stable value about 25m from compensation capacitors.
\subsection{Influence of ballast resistance}
\begin{figure}[!t]
	\centering
	\subfloat[]{\includegraphics[width=3.0in]{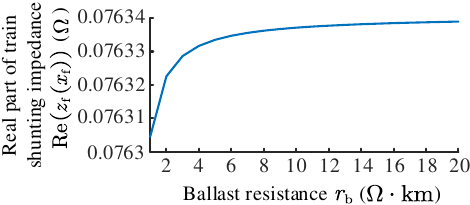}%
		\label{fig_influ_rd_real}}
	\hfil
	\subfloat[]{\includegraphics[width=3.0in]{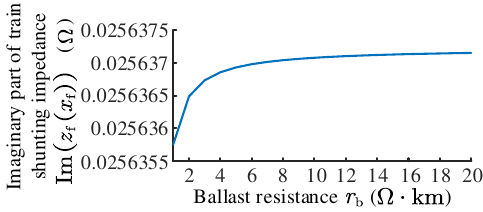}%
		\label{fig_influ_rd_imag}}
	\caption{Relationship between train shunting impedance $z_\text{f}(x_\text{f})$ and ballast resistance $r_\text{b}$. (a) The real part of $z_\text{f}(x_\text{f})$ (b) The imaginary part of $z_\text{f}(x_\text{f})$.}
	\label{fig_influ_rd}
\end{figure}

As a primary parameter of JTC, ballast resistance $r_\text{b}$ is mainly related to the material, thickness and cleanliness of the roadbed, the material and number of sleepers, as well as the weather, temperature and humidity. Based on the simulation conditions of  Fig. \ref{fig_simu_zf}, let $r_\text{b}$ be taken in the range $[1\Omega,20\Omega]$ in steps of 1$\Omega$, and calculate the corresponding train shunting impedance $z_\text{f}(x_\text{f})$. The results are shown in Fig. \ref{fig_influ_rd}.

As can be seen in Fig. \ref{fig_influ_rd}, ballast resistance has a small influence on the real and imaginary part of $z_\text{f}(x_\text{f})$, which are reflected in the 100,000th and millionth digits and the digits thereafter, respectively. In addition, there is a monotonous nonlinear increasing relationship between them. The overall stable value of the real part $\text{Re}(z_\text{f}(x_\text{f}))$ and the imaginary part $\text{Im}(z_\text{f}(x_\text{f}))$ increases with $r_\text{b}$, which is a monotonically increasing relationship, but the increase decreases with $r_\text{b}$ and gradually tends to a steady state. That is, its influence is bounded.
\subsection{Influence of rail impedance}
\begin{figure}[!t]
	\centering
	\subfloat[]{\includegraphics[width=2.8in]{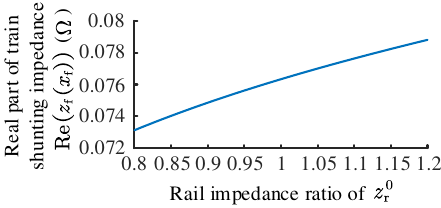}%
		\label{fig_influ_zr_real}}
	\hfil
	\subfloat[]{\includegraphics[width=2.8in]{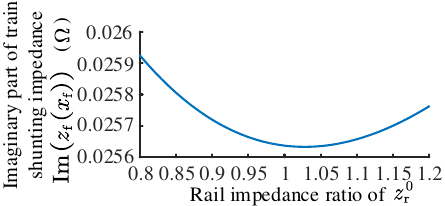}%
		\label{fig_influ_zr_imag}}
	\caption{Relationship between train shunting impedance $z_\text{f}(x_\text{f})$ and rail impedance $z_\text{r}$. (a) The real part of $z_\text{f}(x_\text{f})$ (b) The imaginary part of $z_\text{f}(x_\text{f})$.}
	\label{fig_influ_zr}
\end{figure}

As another primary parameter of JTC, rail impedance $z_\text{r}$ mainly includes two parts, rail resistance and rail inductive impedance. Since these two parts are less affected by external changes, rail impedance $z_\text{r}$ is usually more stable. Here, the changes in train shunting impedance $z_\text{f}(x_\text{f})$ are calculated for $z_\text{r}$ within $\pm 20\%$ of its standard value $z_\text{r}^0$, taken in steps of $\pm 1\%$. The results are shown in Fig. \ref{fig_influ_zr}.

As can be seen in Fig. \ref{fig_influ_zr_real}, $\text{Re}(z_\text{f}(x_\text{f}))$ increases with the increase of $z_\text{r}$, and the relationship between them is approximately linear incremental change. The corresponding linear regression model is
\begin{equation}
	\begin{aligned}
		&\text{Re}(z_\text{f}(x_\text{f}))=a_\text{r}z_\text{r}+b_\text{r} \\
		&z_\text{r}=0.80z_\text{r}^0,0.81z_\text{r}^0,\cdots,1.20z_\text{r}^0
	\end{aligned}
\end{equation}
\noindent where $a_\text{r}$=0.01408, $b_\text{r}$=0.06213 are the parameters of regression model. The corresponding goodness-of-fit is as follows: $\text{SSE}$=$5.4566\times10^{-7}$, $\text{R-square}$=0.995, $\text{RMSE}$=$1.1828\times 10^{-4}$.

As can be seen in Fig. \ref{fig_influ_zr_imag}, there is approximately as quadratic relationship between $\text{Im}(z_\text{f}(x_\text{f}))$ and $z_\text{r}$, whose regression model is
\begin{equation}
	\label{equa_reg_zr_imag}
	\begin{aligned}
		&\text{Im}(z_\text{f}(x_\text{f}))=c_\text{r}(z_\text{r})^2+d_\text{r}z_\text{r}+e_\text{r} \\
		&z_\text{r}=0.80z_\text{r}^0,0.81z_\text{r}^0,\cdots,1.20z_\text{r}^0
	\end{aligned}
\end{equation}
\noindent where $c_\text{r}$=0.005171, $d_\text{r}$=-0.0107 and $e_\text{r}$=0.03117 are the parameters of regression model. The corresponding goodness-of-fit is as follows: $\text{SSE}$=$6.8331\times10^{-10}$, $\text{R-square}$=0.9972, $\text{RMSE}$=$4.2405\times 10^{-6}$. Further, the derivation of (\ref{equa_reg_zr_imag}) yields the corresponding minimum point $(1.034z_\text{r}^0, 0.02564)$. When rail impedance is approximated to its standard value, $\text{Im}(z_\text{f}(x_\text{f}))$ can reach its minimum value.
\section{Simplified model of train shunting impedance based on structural importance of wheel sets}
Based on the above analysis, the main influencing factors of train shunting impedance in terms of JTC are rail impedance and compensation capacitor, and the influence range of compensation capacitor is small. Additionally, the influence of ballast resistance can be approximately ignored. This indicates that the actual range of track line affecting train shunting impedance should be small, and the corresponding number of effective train wheel sets should also be shorter. Therefore, the calculation of train shunting resistance can be simplified, and the corresponding simplified model can be constructed to further clarify the change mechanism.

Based on the structure and formation of the train, the influence degree of each wheel set on train shunting impedance should be different according to their different position in JTC at the same time. From the reliability level, the structural importance of each wheel set needs to be calculated in order to quantitatively analyze the influence degree of each wheel set.

Let $\text{Re}(z_\text{f}^0(x_\text{f}))$ and $\text{Im}(z_\text{f}^0(x_\text{f}))$ denote the overall steady value of the real and imaginary parts of train shunting impedance when the resistance of each wheel set is normal, respectively; while $\text{Re}(z_\text{f}^i(x_\text{f}))$ and $\text{Im}(z_\text{f}^i(x_\text{f}))$ denote the overall steady value of the real and imaginary parts of train shunting impedance when the resistance of $i$-th wheel set is abnormal, respectively. Then the structural importance of $i$-th wheel set based on the real and imaginary of train shunting impedance $p_\text{re}^i$ and $p_\text{im}^i$ can be given by
\begin{equation}
\label{equa_structural_importance_real}
\left\{
\begin{aligned}
	&\Delta p_\text{re}^i = \frac{\left|\text{Re}(z_\text{f}^i(x_\text{f}))-\text{Re}(z_\text{f}^0(x_\text{f}))\right|}{\text{Re}(z_\text{f}^0(x_\text{f}))} \\
	&p_\text{re}^i = \frac{\Delta p_\text{re}^i}{\max(\Delta p_\text{re}^i)}
\end{aligned}
\right.
\end{equation}
\begin{equation}
\label{equa_structural_importance_imag}
	\left\{
	\begin{aligned}
		&\Delta p_\text{im}^i = \frac{\left|\text{Im}(z_\text{f}^i(x_\text{f}))-\text{Im}(z_\text{f}^0(x_\text{f}))\right|}{\text{Im}(z_\text{f}^0(x_\text{f}))} \\
		&p_\text{im}^i = \frac{\Delta p_\text{im}^i}{\max(\Delta p_\text{im}^i)}
	\end{aligned}
	\right.
\end{equation}

\begin{figure}[!t]
	\centering
	\subfloat[]{\includegraphics[width=2.5in]{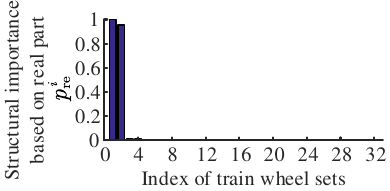}%
		\label{fig_structural_importance_real}}
	\hfil
	\subfloat[]{\includegraphics[width=2.5in]{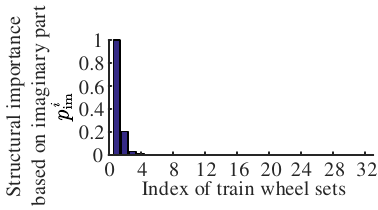}%
		\label{fig_structural_importance_imag}}
	\caption{Structural importance of each wheel set based on train shunting impedance. (a) Structural importance based on the real part of train shunting impedance. (b) Structural importance based on the imaginary part of train shunting impedance.}
	\label{fig_structural_importance}
\end{figure}

Here, obtain the normal value $\text{Re}(z_\text{f}^0(x_\text{f}))$ and $\text{Im}(z_\text{f}^0(x_\text{f}))$ based on the simulation conditions of Fig. \ref{fig_simu_zf}. Let the resistance of the abnormal wheel set is $1\Omega$, and the structural importance of each wheel set can be calculated using (\ref{equa_structural_importance_real}) and (\ref{equa_structural_importance_imag}). The results are shown in Fig. \ref{fig_structural_importance}.

From Fig. \ref{fig_structural_importance_real}, it can be seen that for the real part of train shunting impedance and an 8-car formatted train, the structural importance of the 1st and 2nd wheel set are in the first two places, between which there is not much difference, while the structural importance of the 3rd wheel set decreases rapidly and approaches to 0. In addition, the structural importance of the wheel sets after the 5th wheel set is 0. The above results indicate that the 1st and 2nd wheel set of the train have the greatest influence on the real part of train shunting impedance, whose influence is approximately the same; the 3rd and 4th wheel sets have a smaller influence, which can be approximately ignored; 5th wheel set and the subsequent wheel sets have no influence.

From Fig. \ref{fig_structural_importance_imag}, it can be seen that for the imaginary part of train shunting impedance, the 1st wheel set has the greatest influence, and the ratio of its structural importance to that of the 2nd, 3rd and 4th wheel set is about 4.95, 34.07 and 79.35, respectively; the structural importance of the 5th and 6th wheel set are close to 0, which can be approximately ignored; 7th wheel set and the subsequent wheel sets have no influence.

Based on the above analysis, we can conclude that train shunting impedance cannot be simply modeled using the 1st wheel set. Therefore, for an 8-car formatted train, train shunting impedance should be modeled by the wheel sets of the first car as a simplified model and has nothing to do with the wheel sets of the following 7 cars.
\section{Conclusion}
Based on the transmission line theory, this paper constructs a equivalent six-terminal network model of train shunting impedance by taking the situation of the 8-car formatted train fully occupying jointless track circuit as an example, analyzes the influence of equipment parameters such as resistance of each wheel set, compensation capacitors, ballast resistance, and rail impedance on train shunting impedance on the basis of corresponding simulation and actual data validation, and obtains the following research conclusions:
\begin{enumerate}
\item {For jointless track circuit, train shunting impedance is a complex impedance, which is manifested as a series connection of a shunting resistance and a shunting inductance.}
\item {Train shunting impedance can be expressed as the inverse of the inverse proportional function of wheel set resistance, i.e., there is a nonlinear monotonically increasing relationship between them.}
\item {The influence of compensation capacitor on train shunting impedance is abrupt and bounded.}
\item {The influence of ballast resistance on train shunting impedance is very small and can be ignored.}
\item {The relationship between rail impedance and the real and imaginary parts of train shunting impedance are linear and quadratic function, respectively. And the imaginary part reaches its minimum value when rail impedance is approximated to its standard value.}
\item {Based on the structural importance of train wheel sets, the calculation of train shunting impedance can be simplified to a single car.}
\end{enumerate}

In summary, the study in this paper proposes to replace the existing first train wheel set resistance with train shunting impedance from the contribution of each wheel set to the shunting process, and the constructed model can more accurately describe the train shunting process, which provides theoretical support for the further study of the train shunting problem in JTC.

\bibliographystyle{IEEEtran}
\bibliography{Research_on_Train_Shunting_Impedance_Based_on_Transmission_Line_Theory}










\end{document}